# An arrayed nanoantenna for broadband light emission and detection

**Andrey. E. Miroshnichenko**[*,1], **Ivan S. Maksymov**[1], **Arthur R. Davoyan**[1], **Constantin Simovski**[2,3], **Pavel Belov**[3] **and Yuri S. Kivshar**[1,3]

[1] Nonlinear Physics Centre, Australian National University, Canberra ACT 0200, Australia
[2] Department of Radio Science and Engineering, Aalto University, Aalto 00076, Finland
[3] National Research University of Information Technologies, Mechanics, and Optics (ITMO), St Petersburg 197101, Russia

* Corresponding author: e-mail aem124@physics.anu.edu.au, Phone: +61 2 612 53964, Fax: +61 2 612 8588

We suggest a broadband optical unidirectional arrayed nanoantenna consisting of equally spaced nanorods of gradually varying length. Each nanorod can be driven by near-field quantum emitters radiating at different frequencies or, according to the reciprocity principle, by an incident light at the same frequency. Broadband unidirectional emission and reception characteristics of the nanoantenna open up novel opportunities for subwavelength light manipulation and quantum communication, as well as for enhancing the performance of photoactive devices such as photovoltaic detectors, light-emitting diodes, and solar cells.

**1 Introduction** The field of optical nanoantennas [1, 2] is currently one of the most developing in plasmonics due to their ability to overcome the size and impedance mismatch between subwavelength emitters and free space radiation [3]. Recent studies have demonstrated that the application of radio-frequency (RF) antenna design principles [4] to optical nanoantennas enriches dramatically their functionality and provides novel opportunities for quantum communication [5], subwavelength light enhancement, sensing, medicine and photovoltaic applications [1, 2].

Typically, the radiation of quantum emitters (such as quantum dots [6] or colour centres in nanodiamonds [7]) can be strongly enhanced and highly directed by coupling an emitter to optical Yagi-Uda nanoantennas [8-11]. According to the principle of reciprocity, such kind of antennas can also effectively detect an incoming radiation.

Yagi-Uda like architectures bring considerable advantages over various optical antenna designs [1, 2], which include high antenna gain and directionality originated from the use of inductively and capacitively detuned elements with regard to an active feed [4]. However, in their simplest form Yagi-Uda antennas provide a very narrowband response, with the performance of just a few per cent around their optimal frequency.

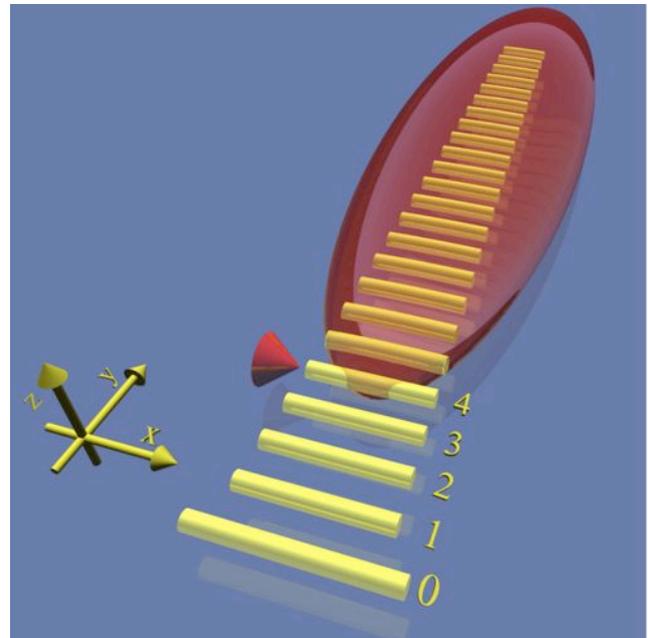

**Figure 1** Optical unidirectional nanoantenna consisting of silver nanorods of gradually varying length. Red arrow denotes one of the positions and the polarization of the emitter.

In RF region, the bandwidth can be extended by increasing the diameter of antenna elements, which is usually negligibly small as compared with the operating wavelength [4]. The relative diameter of optical nanoantenna el-elements is effectively larger than in RF due to fabrication constraints [1].

Nevertheless, this problem can also be addressed by referring to RF antenna technology and adopting the design of directional array antennas such as *log-periodic antennas*. Owing to a gradually varying length of their elements, where each of them can be driven by an external radiation, log-periodic antennas offer broadband and high directivity operation compared to Yagi-Uda-like antennas [4].

In this Letter, we study theoretically an optical broadband unidirectional arrayed nanoantenna consisting of plasmonic nanorods of gradually varying length. In analogy to RF log-periodic antennas, the nanorods can be driven by quantum emitters placed in the near-field zone. As the optical response of the antenna depends on the resonance frequency of the excited element given by its dimensions, the excitation of shorter nanorods shifts the antenna operation towards shorter wavelengths. Therefore, an effective blue-shift of operational wavelength is expected with the same antenna design by properly choosing the position and wavelength of the excited emitter.

The investigated architecture is considerably simpler than that of the most common RF log-periodic antennas with decreasing spacing of the elements [4]. But at the same time it offers same advantages over Yagi-Uda antennas such as a wide operating bandwidth and can be fabricated by using current nanotechnology techniques (e.g. [1]).

**2 Design** We consider an optical nanoantenna (Fig. 1) consisting of an array of equally spaced silver nanorods (for the sake of computational time located in free space since any practical design is not the goal of the present Letter). The design and interpretation of the gradually varying length of the nanorods is carried out by using the effective wavelength rescaling approach, which accounts for inevitable losses in metal at optical wavelengths [1]. The permittivity of silver used in our study is based on a Drude model fit based on the published optical constants data [12].

We use commercial CST Microwave Studio software to calculate the far-field emission diagrams and the antenna radiation efficiency that measures the losses that occur throughout the antenna while it is operating at a given wavelength. The radiation efficiency can be expressed as the ratio of the power radiated by the antenna ($P_{rad}$) in the *y*-direction (Fig. 1) to the input power ($P_{in}$) delivered by the emitter to the antenna [4]

$$\eta = P_{rad}/P_{in} = P_{rad}/(P_{rad} + P_{loss}), \quad (1)$$

where $P_{loss}$ denotes the losses caused by the propagation losses of the antenna mode and absorption due to the field induced directly in the metal by the emitters [13].

The impact of gradually varying elements is stronger for longer antennas. In our study we consider 42 nanorods. This value is chosen such that the fabrication of the smallest nanorod would be affordable using current fabrication technologies offering an accuracy of 10-20 nm [1]. We compared our results for antennas with a larger and smaller number of elements, and all of them exhibit similar and consistent results.

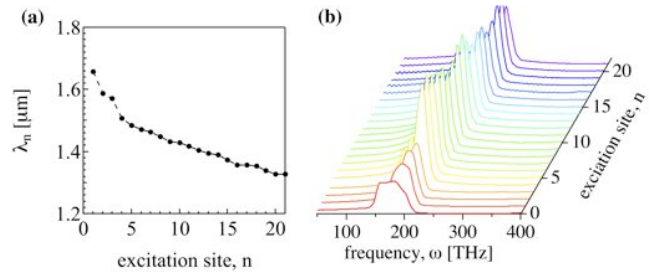

**Figure 2** (a) Operating wavelength and (b) near-field radiated power spectra of the antenna excited alternately with point broadband radiation sources placed at the excitation sites $n = 1…20$.

By means of numerical simulation we define the optimal dimensions of nanorods that offer the best efficiency for the chosen 42-element geometry. We set the diameter of all nanorods to $d = 50$ nm and the edge-to-edge inter-nanorod spacing as $w = 30$ nm. The edge nanorod $n = 0$ of length 485 nm is not used for excitation and can be considered as an effective reflector. We notice that the use of the term "reflector" is justified by the same criteria of inductive and capacitive antenna elements found in Yagi-Uda like antennas (see discussion below). The length variation is applied to all other nanorods with $n = 1...41$. The overall length variation from nanorod $n = 1$ to nanorod $n = 41$ constitutes 390 nm and 270 nm, respectively.

**3 Results and Discussion** The principle of reciprocity is used to investigate *near-field* spectral characteristics of the nanoantenna operating in emission and reception regimes. First, we excite the nanoantenna with a point-like broadband emitter placed near the feeding nanorods $n = 1...20$ and probe the power radiated in the *y*-direction (Fig. 1). In all simulations, we modelled the emitter as a generic dipole polarized along the *x*-direction and placed 5 nm away from the nanorods. Then, we illuminate the front-end of antenna with a broadband incident plane wave pulse and probe the field with pointlike $E_x$-field detectors placed at the excitation sites $n = 1...20$ at 5 nm from the nanorods, thus reproducing an experimental scheme used for near-field measurements of a optical analogue of Yagi-Uda antenna [14].

In the emission regime, the individual antenna elements are excited at resonant wavelengths $\lambda_n$ [Fig. 2(a)] corresponding to the maxima of the power spectra in Fig. 2(b). One can see that the optimal excitation wavelength $\lambda_n$ is blue-shifted for shorter nanorods. For instance, the maximum of the emitted power is observed at 1.65 μm for $n = 1$ and shits towards 1.32 μm for $n = 20$, thus, cover-ing more than 300 nm operating bandwidth, which is much wider as compared to the classical Yagi-Uda design [4].

In the reception regime, the broadband performance of the antenna manifests itself by the dependence of the local field enhancement at the nanorods at the resonant wavelengths. The calculated near-field spectra of the detectors agree with the results in Fig. 2, and cover the same 300 nm operational bandwidth from 1.32 to 1.65 µm.

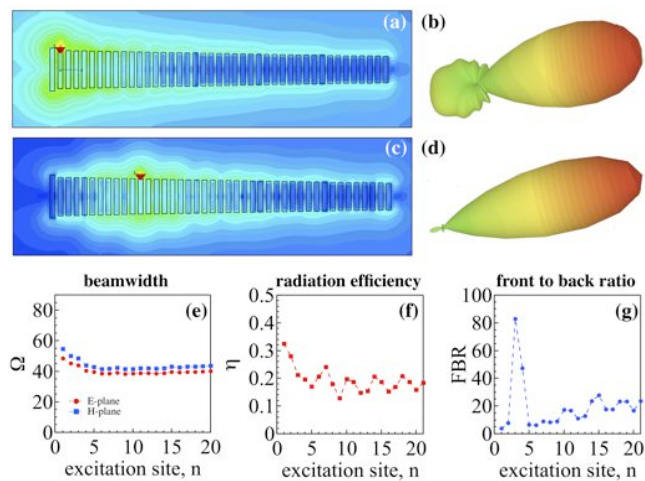

**Figure 3** (a,c) |$E$|-field profile and (b,d) the far-field angular emission diagram of the antenna excited via the 1$^{st}$ (a,b) and 10$^{th}$ (c,d) nanorods. (e) Beamwidth $\Omega$, (f) radiation efficiency $\eta$, and (g) front to back ratio of the nanoantenna as a function of the excitation site number $n$, respectively.

Figures 3(a, c) show the electric near-field profiles for the antenna excited at the 1st and 10th nanorods at the wavelengths specified in Fig. 2(a). This result demonstrates that the excitation at a shorter nanorod results in a shorter effective length of the nanoantenna $L_{eff}$. By analysing the far-field emission diagrams [Figs. 3(b, d)], we plot the beamwidth $\Omega$ [Fig. 3(e)], the antenna radiation efficiency $\eta$ [Fig. 3(f)], and front to back ration (FBR) [Fig. 3(g)], as a function of the excitation cite $n$. As the directive gain $D$ taken along the $y$-direction is inversely proportional to $\Omega$, we notice a good correlation between the *near-field* power spectra [Fig. 2(b)] and the *far-field* result in Fig. 3(e).

By artificially neglecting the losses in calculations, we studied the impact of the absorption losses in metal on the overall performance of the nanoantenna. The excitation of shorter nanorods at corresponding lower wavelengths $\lambda_n$ [Fig. 2(a)] effectively increases the intrinsic absorption losses of real metal and the propagation losses of the antenna mode. The passive nanorods between the reflector and the excited nanorod perform similarly to a reflector of Yagi-Uda like antennas for excitation at $n = 1...5$, which is clearly indicated by a strong FBR [see Fig. 3(g)]. The excitation at $n > 5$ results in larger absorption [15] and thus impairs $\eta$ [Fig. 3(f)]. We also notice that $L_{eff}$ decreases with $n$, which leads to an inevitable increase in $\Omega$ and a decrease in $D$. However, the propagation losses through the antenna decrease as one decreases $L_{eff}$.

The combination of these effects maintains $\Omega$ at nearly the same level for $n = 6...20$. It offers equal conditions for the collection of the light emitted by the antenna at the corresponding $\lambda_n$ within a certain cone given by the numerical aperture of detection optics. The principle of reciprocity guarantees the same conditions for the antenna reception. $\eta$ averaged along the operating band is high enough taking into account a wide operating band and a high directivity of the antenna. One can enhance the performance of light-emitting diodes, solar cells etc. by placing a single nanoantenna close to an active region of a device. The antenna can be also used for the generation of hot electron-hole pairs arising from plasmon decay at different nanorods, resulting in the broadband detection of light [16].

**In conclusion**, we have demonstrated an optical directional arrayed nanoantenna that offers novel opportunities for optimization of both radiation and local field enhancement. Employed as a receiver, it can enhance the performance of photoactive devices (photovoltaic detectors, light-emitting diodes, etc) in a broad frequency range. Selective excitation or probing of individual elements of the optical nanoantenna not only brings the rich functionality of the family of broadband RF array antennas to the nanoscale, but also dramatically simplifies their construction making them feasible using current nanotechnologies.

**Acknowledgements** This work was supported by the Australian Research Council and the Ministry of Education and Science in Russia.